# Moist adiabats with multiple condensing species: A new theory with application to giant planet atmospheres

Cheng Li[1,2,*], Andrew P. Ingersoll[1], Fabiano Oyafuso[2]

[1]Division of Geological and Planetary Science, California Institute of Technology, Pasadena, CA, 91125

[2]Jet Propulsion Laboratory, California Institute of Technology

* Corresponding author: cli@gps.caltech.edu

# Abstract

We derived a new formula for calculating the moist adiabatic temperature profile of an atmosphere consisting of ideal gases with multiple condensing species. This expression unifies various formulas published in the literature and can be generalized to account for chemical reactions. Unlike previous methods, it converges to machine precision independent of mesh size. It accounts for any ratio of condensable vapors to dry gas, from zero to infinity, and for variable heat capacities as a function of temperature. Because the derivation is generic, the new formula is not only applicable to planetary atmosphere in the solar system, but also to hot Jupiters and brown dwarfs in which a variety of alkali metals, silicates and exotic materials condense. We demonstrate that even though the vapors are ideal gases, they interact in their effects on the moist adiabatic lapse rate. Finally, we apply the new thermodynamic model to study the effects of downdrafts on the distribution of minor constituents and thermal profile in the Galileo probe hotspot. We find that the Galileo Probe measurements can be interpreted as a strong downdraft that displaces an air parcel from 1 bar to the 4 bar level.

# 1. Introduction

The starting point of modeling atmospheric thermal structure for giant planets is to calculate a moist adiabatic lapse rate. Such process was initially discussed by Weidenschilling; Lewis (1973) and was later further developed by Atreya (1987). Their expressions for the moist adiabatic lapse rate are used in the numerous literature onwards. However, little attention has been paid to the various assumptions made in the calculation of the temperature profile, because the troposphere of a giant planet is rarely observed except for a few holes in the clouds like the 5-um hot spot on Jupiter. Previous observations are not able to distinguish various thermodynamic models.

Recently, the Juno spacecraft has made several flybys to Jupiter, and the Microwave Radiometer (MWR) onboard the spacecraft has measured the thermal emission from Jupiter's atmosphere from the cloud top at about 1 bar to a hundred bars. Because the measured absolute brightness temperature is precise to about a few percent, and the limb darkening is precise to one tenth of a percent (Janssen et al. 2016), traditional Jovian thermodynamics – assuming constant heat capacity and small mixing ratios of condensates – needs to be carefully reviewed and refined according to the requirement of the new instrument. Furthermore, numerous exotic exoplanets exhibit the condensation of alkali metals, silicates, irons, etc in their atmospheres (Morley et al. 2012). The traditional moist adiabatic theory for single condensing species deserves a revisit. Several authors have improved the original model of Weidenschilling; Lewis (1973) using different methods and assumptions. For example, Atreya; Romani (1985) considered ammonia solution and chemical reactions forming $NH_4SH$, but they assumed a diluted atmosphere, meaning that the mixing ratios of condensable species are small. Pierrehumbert (2010) and Leconte et al.

(2016) derived the formula that accounts for large mixing ratios, but it is limited to one condensable species. Here we derive a generic reversible moist adiabatic lapse rate for an atmosphere consisting of idea gases with multiple condensable species. The formula is:

$$\Gamma_m = \frac{d \ln T}{d \ln P} = \frac{1 + \sum_i \eta_i \beta_i}{\widehat{c_p}/R + \frac{\sum_i \eta_i \beta_i^2 + (\sum_i \eta_i \beta_i)^2}{1 + \sum_i \eta_i}}$$

$$\widehat{c_p} = \frac{\overline{c_p}}{x_d + \sum_i x_{v_i}} = \frac{x_d c_{pd} + \sum_i x_{v_i} c_{pv_i} + \sum_i x_{c_i} c_{pc_i}}{x_d + \sum_i x_{v_i}},$$

(1)

where superscripts $d, v_i$ and $c_i$ stand for dry air (comprised of all non-condensable gases), $i$-th vapor, and its cloud respectively. $x_d, x_{v_i}, x_{c_i}$ are their molar abundances in one mole of heterogeneous gas mixture. This definition of $x_d$ is the same as the volume mixing ratio if no cloud component is present. When condensation happens, part of $x_{v_i}$ is converted to $x_{c_i}$ following the saturation vapor curve but $x_d$ keeps constant throughout the atmosphere. For example, in the Jovian atmosphere, $x_d$ represents the ratio of $H_2$/He molecules to all molecules in the air parcel, including condensed clouds. $x_{v_1}$ and $x_{v_2}$ are the mixing ratios of ammonia and water gas; $x_{c_1}$ and $x_{c_2}$ are the mixing ratios of ammonia and water clouds. The volume mixing ratio of ammonia gas is $x_{v_1}/(x_d + x_{v_1} + x_{v_2})$. $c_{pd}, c_{pv_i}, c_{pc_i}$ are the molar heat capacities evaluated at the current temperature. $\eta_i = x_{v_i}/x_d$ is the gas-phase molar mixing ratio with respect to dry air. In an atmosphere with no dry component, $\eta_i$ should be treated as the limiting value in which $x_d \to 0$. The example of a steam atmosphere is given below. $\beta_i = L_i/(RT)$ is the ratio of latent heat over $RT$ for each condensable species.

Equation (1) unifies various formulas published in the literature. For example, it reduces to the moist adiabatic lapse rate derived by Pierrehumbert (2010) and Leconte et al. (2016) for an atmosphere with a single condensable species. Assuming that the vapor mixing ratio is small, equation (1) reduces to that derived by Weidenschilling; Lewis (1973) and Atreya (1987) without chemical reactions. Furthermore, equation (1) reduces to the Clausius-Clapeyron relation for a single component steam atmosphere ($\eta \to \infty$):

$$\frac{d \ln T}{d \ln P} = \lim_{\eta \to \infty} \frac{1 + \eta \beta}{\frac{\widehat{c_p}}{R} + \frac{\eta \beta^2 + \eta^2 \beta^2}{1 + \eta}}$$

$$= \lim_{\eta \to \infty} \frac{\eta \beta}{\frac{\widehat{c_p}}{R} + \eta \beta^2} = \frac{1}{\beta} = \frac{RT}{L}. \qquad (2)$$

Equation (1) also makes clear a fact which appears not to have been appreciated before, namely that for an atmosphere with multiple condensable species, these is a quadratic term $(\sum_i \eta_i \beta_i)^2$, which is a square of a sum over all condensable species, in the expression of the moist adiabatic lapse rate.

The remaining paper is organized into three sections. Section 2 derives equation (1). The methods to include chemical reactions is discussed afterwards. Section 3 compares the numerical solution against the traditional method used in (Atreya 1987). Section 4 applies the thermodynamic model to interpret the Galileo Probe result. Section 5 concludes.

## 2. Thermodynamic model

*2.1 moist adiabatic model without chemical reactions*

The moist adiabatic temperature profile of an atmosphere consisting of an ideal gas mixture can be calculated in two ways. One way is to use the differential form of the first law of thermodynamics as described in (Leconte et al. 2016; Pierrehumbert 2010; Weidenschilling; Lewis 1973), in which the thermal profile is obtained by integrating the lapse rate, while keeping track of the condensed species. This method is faster to calculate, but its accuracy depends on the vertical resolution, especially on the ability to locate the cloud bottom because the lapse rate is discontinuous where a phase transition occurs, i.e. when vapor condenses into liquid, or when liquid freezes. Large numerical errors can occur if the cloud base is off by one grid point. In order to calculate the moist adiabatic profile accurately, one has to insert numerical grids to represent the cloud base adaptively. The drawback is avoided using an alternative approach, which is to use the integral form of the first law of thermodynamics as described in Emanuel (1994). The entropy of the air parcel is explicitly evaluated, and temperature is determined at each level to conserve entropy during adiabatic displacement. Thus, the resulting thermal profile is independent of numerical mesh size.

In the following paragraphs, we consider the process that a heterogeneous air parcel goes through reversible and adiabatic expansion. "Heterogeneous" means that the air parcel contains both gas phase and condensed phase materials. "Reversible" means that all condensates remain in the parcel and don't precipitate. "Adiabatic" means without exchange of heat. We first derive the formula of specific entropy (potential temperature). The derivation follows chapter 4 in Emanuel (1994), but extends to an ideal atmosphere with varying heat capacity as a function of temperature and with multiple condensable species. Then, we take the derivative of the entropy function to obtain the moist adiabatic lapse rate, equation (1).

To start with, we assume that an air parcel is an ideal mixture of dry air, vapors and clouds (condensed liquid and solid). The derivation in this section uses the following properties of an ideal mixture:

1) Each gaseous component satisfies the ideal gas law $PV = nRT$.
2) Total pressure is the sum of the partial pressure of each gaseous component.
3) Total entropy is the sum of the entropy of each component.
4) Heat capacity and latent heat are functions of temperature only.
5) The specific volume of a condensed component is neglected.

Using the first law of thermodynamics for one mole of pure ideal gas:

$$T dS = dH - V dP$$
$$= dH - RT d\ln P$$
$$= c_p(T) dT - RT d\ln P . \tag{3}$$

The entropy at temperature $T$ and pressure $P$ is obtained from integrating equation (3):

$$S(T,P) = \int c_p(T) \frac{dT}{T} - R d\ln P = s(T) - RT \ln P . \tag{4}$$

We have neglected the integration constant, which represents the entropy at any reference temperature and pressure, and $s(T)$ denotes the result of the integral $\int c_p(T) \frac{dT}{T}$. For a gas with constant heat capacity $c_p$, one obtains $s(T) = c_p \ln T$. Otherwise, $s(T)$ is taken as a known function which comes from a chemical library such as NIST (Linstrom; Mallard 2001). The entropy of a condensed component is related to the gaseous component through the definition of latent heat:

$$L(T) = T \times [S_v(T,P) - S_c(T,P)]$$

$$S_c(T,P) = S_v(T,P) - \frac{L(T)}{T} \quad (5)$$

where $S_v(T,P)$ is the molar entropy of saturated vapor. The first derivative of latent heat is given by Kirchhoff's equation (Zemansky 1968):

$$\frac{d\,L(T)}{d\,T} = c_p(T) - c_c(T) \quad (6)$$

where $c_c(T)$ is the heat capacity of the condensed phase.

Consider one mole of gas-cloud mixture, in which the molar amount of dry air is $x_d$; the molar amounts of $i$-th vapor and cloud are $x_{v_i}$ and $x_{c_i}$ respectively. The total molar amount of condensable componet $i$ is conserved during phase change, which is denoted as $x_{t_i} = x_{v_i} + x_{c_i}$ and $x_d + \sum_i x_{t_i} = 1$. The entropy of the heterogeneous air parcel is the sum of the entropy of each homogeneous component:

$$\begin{aligned} S(T,P) &= x_d S_d(T,P) + \sum_i x_{v_i} S_{v_i}(T,P) + \sum_i x_{c_i} S_{c_i}(T,P) \\ &= x_d s_d(T) + \sum_i x_{t_i} s_{v_i}(T) - R\left(x_d \ln P_d + \sum_i x_{t_i} \ln P_{v_i}\right) - \sum_i x_{c_i} \frac{L_i(T)}{T} \\ &= \bar{s}(T) - R\left(x_d \ln P_d + \sum_i x_{t_i} \ln P_{v_i}\right) - \sum_i x_{c_i} \frac{L_i(T)}{T} \end{aligned} \quad (7)$$

where $P_d$ and $P_{v_i}$ are the partial pressures defined by:

$$P_d = \frac{x_d}{x_d + \Sigma_i x_{v_i}} P, \quad P_{v_i} = \frac{x_{v_i}}{x_d + \Sigma_i x_{v_i}} P \tag{8}$$

and $\bar{s}(T)$ is defined as:

$$\bar{s}(T) = \int \left( x_d c_{pd}(T) + \sum_i x_{t_i} c_{pv_i}(T) \right) \frac{dT}{T}$$

$$= x_d s_d(T) + \sum_i x_{t_i} s_{v_i}(T) \tag{9}$$

Conventionally, entropy is expressed in terms of potential temperature ($\theta$), which is the temperature of an air parcel when it is adiabatically displaced to a reference pressure ($P^0$), at which all condensates evaporate (below the base of the deepest cloud).

$$S(\theta, P^0) = \bar{s}(\theta) - R \left( x_d \ln P_d^0 + \sum_i x_{t_i} \ln P_{v_i}^0 \right) \tag{10}$$

Let $S(T, P) = S(\theta, P^0)$ gives an implicit function for potential temperature:

$$\bar{s}(\theta) = \bar{s}(T) + R \ln \left( \frac{P_d^0}{P_d} \right) - R \sum_i x_{t_i} \ln \frac{x_{v_i}}{x_{t_i}} - \sum_i \frac{L_i(T)}{T} x_{c_i}$$

$$= x_d s_d(T) + \sum_i x_{t_i} s_{v_i}(T) + R \ln \left( \frac{P_d^0}{P_d} \right) - R \sum_i x_{t_i} \ln \frac{x_{v_i}}{x_{t_i}} - \sum_i \frac{L_i(T)}{T} x_{c_i} \tag{11}$$

The potential temperature defined in equation (11) is known as "liquid water potential temperature" in Emanuel (1994) (equation 4.5.15), but it has been generalized for gases with non-constant heat capacities. We also define the dry potential temperature $\theta_d$ by dropping the latent heat term in equation (11):

$$\bar{s}(\theta_d) = \bar{s}(T) + R \ln\left(\frac{P_d^0}{P_d}\right) - R \sum_i x_{t_i} \ln \frac{x_{v_i}}{x_{t_i}}, \tag{12}$$

to describe the potential temperature of an air parcel without condensates when it is adiabatically compressed to a reference pressure $P^0$. Note that equation (11) and (12) reduce to the nominally defined potential temperature if no vapor has condensed and if the heat capacity is constant, as can be seen by setting $x_{c_i} = 0$, $x_{v_i} = x_{t_i}$, $\bar{s}(T) = c_p \ln T$ and $\bar{s}(\theta) = c_p \ln \theta$

$$\theta = \theta_d = T\left(\frac{P_d^0}{P_d}\right)^{R/c_p} = T\left(\frac{P^0}{P}\right)^{R/c_p} \tag{13}$$

The moist adiabatic lapse rate $\Gamma_m = \frac{d \ln T}{d \ln P}$ is derived by taking the derivative of equation (11) with respect to $\ln P$. The left-hand side is zero and the right-hand side has four terms:

The first term is:

$$\frac{d\bar{s}(T)}{d \ln P} = T \frac{d\bar{s}(T)}{dT} \frac{d \ln T}{d \ln P} = \left(x_d c_{pd}(T) + \sum_i x_{t_i} c_{v_i}(T)\right) \Gamma_m \tag{14}$$

The second term is:

$$R \frac{d(\ln P_d^0 - \ln P_d)}{d \ln P} = -R \frac{d \ln P_d}{d \ln P} \tag{15}$$

The third term is:

$$-R \sum_i x_{t_i} \frac{d(\ln x_{v_i} - \ln x_{t_i})}{d \ln P} = -R \sum_i x_{t_i} \frac{d \ln x_{v_i}}{d \ln P} \tag{16}$$

The fourth term is:

$$-\sum_i \frac{d}{d \ln P}\left(\frac{L_i(T)}{T} x_{c_i}\right)$$

$$= -\sum_i \left(\frac{x_{c_i}}{T}\frac{dL_i(T)}{d \ln P} + \frac{L_i(T)}{T}\frac{dx_{c_i}}{d \ln P} - \frac{x_{c_i}L(T)}{T}\Gamma_m\right)$$

$$= \sum_i \frac{x_{v_i}L_i(T)}{T}\frac{d \ln x_{v_i}}{d \ln P} + \sum_i x_{c_i}\left(\frac{L_i(T)}{T} - \left(c_{pv_i}(T) - c_{c_i}(T)\right)\right)\Gamma_m \qquad (17)$$

Note that we have used equation (6) and $x_{t_i} = x_{v_i} + x_{c_i}$ to derive equation (17). These four terms are expressed in three gradients: $\Gamma_m$, $d \ln P_d / d \ln P$ and $d \ln x_{v_i} / d \ln P$. The first one is what we want and the last two are unknown. Equation (8) is used to derive the expression for the last two gradients, as follows: First, take the logarithm of the second equation of (8) and then take differentials:

$$d \ln \frac{P_{v_i}}{P} = d \ln x_{v_i} - d \ln\left(x_d + \sum_i x_{v_i}\right) = d \ln x_{v_i} - \frac{\sum_i dx_{v_i}}{x_d + \sum_i x_{v_i}}. \qquad (18)$$

For a system with a single condensable component, equation (18) is trivial to solve for $d \ln x_{v_i}$ in terms of $d \ln P_{v_i}$. But for a system with multiple condensable components, a set of equations needs to be solved simultaneously. The way to solve this set of equations is to solve for the case of two species first and then generalize the solution for multiple species. The solution is:

$$\frac{d \ln x_i}{d \ln P} = \frac{d \ln P_{v_i}}{d \ln P} - 1 + \sum_j \eta_j \left(\frac{d \ln P_{v_j}}{d \ln P} - 1\right), \qquad (19)$$

where $\eta_i = x_{v_i}/x_d$ is the molar mixing ratio with respect to dry air. Second, the vapor pressure is proportional to the molar mixing ratio:

$$\frac{d\ln P_d}{d\ln P} = \frac{d\ln P_{v_i}}{d\ln P} - \frac{d\ln x_{v_i}}{d\ln P} = 1 - \sum_j \eta_j \left(\frac{d\ln P_{v_j}}{d\ln P} - 1\right) \tag{20}$$

$x_d$ vanishes in equation (20) because it is a constant. Equation (19) and (20) can be further simplified using the ideal gas form of the Clausius-Clapeyron relation:

$$\frac{d\ln P_{v_i}}{d\ln T} = \frac{L_i(T)}{RT} = \beta_i(T) \tag{21}$$

Substitute equation (21) into equation (19) and (20) gives:

$$\frac{d\ln P_d}{d\ln P} = 1 - \sum_j \eta_j (\beta_j \Gamma_m - 1) \tag{22}$$

$$\frac{d\ln x_i}{d\ln P} = \beta_i \Gamma_m - 1 + \sum_j \eta_j (\beta_j \Gamma_m - 1) \tag{23}$$

We have omitted (T) in $\beta(T)$ and $c_p(T)$ for clarity. Substitute equation (22) and (23) into equations (14) – (17) gives,

$$\frac{d\bar{s}(T)}{d\ln P} = \left(x_d c_{pd} + \sum_i x_{t_i} c_{pv_i}\right) \Gamma_m \tag{24}$$

$$R \frac{d(\ln P_d^0 - \ln P_d)}{d\ln P} = -R\left(1 + \sum_j \eta_j - \sum_j \eta_j \beta_j \Gamma_m\right) \tag{25}$$

$$-R \sum_i x_{t_i} \frac{d\left(\ln\left(\frac{x_{v_i}}{x_{t_i}}\right)\right)}{d\ln P} = -R \sum_i x_{t_i} \left(\beta_i \Gamma_m - 1 - \sum_j \eta_j + \sum_j \eta_j \beta_j \Gamma_m\right) \tag{26}$$

$$-\sum_i \frac{d}{d\ln P}\left(\frac{L_i(T)}{T} x_{c_i}\right) = \tag{27}$$

$$R \sum_i \beta_i x_{v_i} \left(\beta_i \Gamma_m - 1 - \sum_j \eta_j + \sum_j \eta_j \beta_j \Gamma_m\right) + \sum_i x_{c_i} \left(R\beta_i - (c_{pv_i} - c_{c_i})\right) \Gamma_m$$

Equations (24) – (27) sum to zero. Collecting all terms involving $\Gamma_m$ to the left-hand side and all other terms to the right-hand side results:

$$\Gamma_m = \frac{d \ln T}{d \ln P} = \frac{1 + \sum_i \eta_i \beta_i}{\hat{c}_p/R + \frac{\sum_i \eta_i \beta_i^2 + (\sum_i \eta_i \beta_i)^2}{1 + \sum_i \eta_i}}$$

$$\hat{c}_p = \frac{\bar{c}_p}{x_d + \sum_i x_{v_i}} = \frac{x_d c_{pd} + \sum_i x_{v_i} c_{pv_i} + \sum_i x_{c_i} c_{pc_i}}{x_d + \sum_i x_{v_i}}$$

(28)

As far as the authors know, the quadratic term $(\sum_i \eta_i \beta_i)^2$ has not been mentioned in the previous literature. For example, there are two condensable species in Jovian atmosphere, ammonia and water. Therefore,

$$\left(\sum_i \eta_i \beta_i\right)^2 = \left(\eta_{\mathrm{NH_3}} \beta_{\mathrm{NH_3}} + \eta_{\mathrm{H_2O}} \beta_{\mathrm{H_2O}}\right)^2$$
$$= \left(\eta_{\mathrm{NH_3}} \beta_{\mathrm{NH_3}}\right)^2 + \left(\eta_{\mathrm{H_2O}} \beta_{\mathrm{H_2O}}\right)^2 + 2\left(\eta_{\mathrm{NH_3}} \beta_{\mathrm{NH_3}}\right)\left(\eta_{\mathrm{H_2O}} \beta_{\mathrm{H_2O}}\right)$$

(29)

The cross term, $2(\eta_{\mathrm{NH_3}} \beta_{\mathrm{NH_3}})(\eta_{\mathrm{H_2O}} \beta_{\mathrm{H_2O}})$, represents an interaction between ammonia and water when both of them are condensing. It is negligible in the Jovian troposphere because the ammonia cloud and the water cloud are quite separated. However, in brown dwarfs' atmospheres or in the deep atmosphere of Jovian planets where the temperature is around 1000 K, ZnS and KCl condense almost simultaneously. The cross term has same magnitude as the other terms in equation (28) and cannot be neglected.

## 2.2  moist adiabatic model with chemical reactions

The previous analysis focused on analyzing an ideal moist adiabat without chemical reactions. In reality, chemical reactions involving two compounds to form a third compound are common and they alter the thermal and compositional profile. For example, in Jupiter's atmosphere, $NH_3$ and $H_2S$ react to form solid $NH_4SH$ cloud when the product of their partial pressures exceeds an equilibrium constant ($K$), which is given in Lewis (1969):

$$\log \frac{K}{\text{atm}^2} = 14.82 - \frac{4705 \text{ K}}{T} \tag{30}$$

This reaction is predicted by a chemical equilibrium model and was confirmed in the laboratory experiment (Magnusson 1907). Yet, whether this cloud layer exists in the presence of complicated dynamics is still controversial. Adding this reaction into our previous entropy function is straightforward. Similar to equation (5), we wrote the entropy of $NH_4SH$ solid as:

$$S_{NH_4SH}(T,P) = S_{NH_3}(T,P) + S_{H_2S}(T,P) - \frac{L_{NH_4SH}(T)}{T} \tag{31}$$

The entropy of the $NH_4SH$ cloud is thus:

$$\begin{aligned}
&x_{NH_4SH}\left(S_{NH_3}(T,P) + S_{H_2S}(T,P) - \frac{L_{NH_4SH}(T)}{T}\right) \\
&= x_{NH_4SH} S_{NH_3}(T,P) + x_{NH_4SH} S_{H_2S}(T,P) - x_{NH_4SH} \frac{L_{NH_4SH}(T)}{T}
\end{aligned} \tag{32}$$

Define $x_{t_{NH3}} = x_{v_{NH_3}} + x_{c_{NH_3}} + x_{NH_4SH}$ and $x_{t_{H2S}} = x_{v_{H_2S}} + x_{c_{H_2S}} + x_{NH_4SH}$ as the total molar mixing ratio of $NH_3$ and $H_2S$, the entropy function is the same as equation (7) and the potential temperature is the same as that defined in equation (11).

*2.3    Neutrally stable atmospheric profile*

The derived reversible moist adiabatic model is neutrally stable in the vertical because all condensates are suspended in the air instead of raining out. Observations (Xu; Emanuel 1989) and numerical models (Bretherton; Smolarkiewicz 1989) show that the temperature profile in the Earth tropics follows that of a reversible moist adiabat even though the majority of the tropics is unsaturated, a consequence of compensating subsidence induced by spreading gravity waves. Thus, the virtual temperature profile, i.e. the density profile, obtained by lifting an air parcel adiabatically and reversibly may be the same as the virtual temperature profile in the unsaturated part of the atmosphere. To solve for an atmospheric profile that is neutrally stable considering the molecular weight effect can be simply done by solving a profile of constant virtual potential temperature:

$$\theta_v = \theta \frac{x_d + \sum_i x_{v_i}}{x_d + \sum_i \epsilon_i x_{t_i}}, \tag{33}$$

where $\epsilon_i = \mu_i/\mu_d$ is the ratio of the molecular weight of a condensable species to the dry air, and the potential temperature $\theta$ is implicitly defined in equation (11). The numerical method is laid out in the next section.

## 3.    Numerical method and model comparison

We construct a numerical model that solves for the moist adiabatic temperature profile using equation (11). Because equation (11) is an implicit function of temperature, an iterative method shall be used. At any pressure level, the iteration starts from an initial guess of temperature. Then the saturation vapor pressure of a condensable species is calculated. If it is smaller than the partial vapor pressure, the species condenses either to a liquid or to a solid depending on the temperature. The condensing process is done sequentially for all condensable species to reach an equilibrium state. This process has to be repeated for several times because condensation of one species will change the partial pressure of the others. After that, entropy is computed for the equilibrium state. If the entropy is not the same as the required entropy, another iteration begins with an updated temperature calculated by the secant method. The iteration usually converges in a few iterations.

Special consideration needs to be applied at the triple point, because the above method only applies to a pure liquid phase or solid phase. At the triple point of one substance, the temperature gradient is zero. Liquid phase coexists with solid phase to keep a constant temperature and partial pressure. In the fusion process, entropy takes a finite jump between those two states although temperature maintains. If the required entropy is in the middle of the above two situations, the secant method will stop at the fusion temperature ($T_{tr}$) with either pure liquid or solid. A practical and elegant way to handle the triple point equilibrium is to calculate two equilibrium states at $T_1 = T_{tr} + \Delta T$ and $T_2 = T_{tr} - \Delta T$ representing a pure liquid phase and a pure solid phase. $\Delta T = 10^{-8}$ is an arbitrary small number. Because entropy is a linear function of mixing ratio during fusion, the equilibrium state at the triple point is given by a linear interpolation between the liquid state and the solid state:

$$x = \frac{s_2 - s_0}{s_2 - s_1} x_1 + \frac{s_0 - s_1}{s_2 - s_1} x_2 \tag{34}$$

where $x$ is the molar mixing ratio; $s_1$ and $s_2$ are the entropies at two states; $s_0$ is the required entropy.

We construct a nominal Jupiter's atmosphere using the iteration of equation (11), which is named as "iteration profile", and then check whether the analytical lapse rate derived in equation (28) goes through the numerical one. The reason to do such a comparison is to illustrate that there are certain regions in the atmosphere where the analytical expression fails to apply, for example, near the $NH_4SH$ cloud and near the triple point of water. The gases included in the atmosphere are $H_2$, He, $CH_4$, $NH_3$, $H_2S$ and $H_2O$. Their solar abundances are according to Asplund et al. (2009), standard enrichment factors are 1.0, 0.81, 3.9, 5.0, 3.0, 5.0 with respect to $H_2$. The heat capacity of hydrogen is a function of temperature, which depends upon the ratio of ortho-hydrogen to para-hydrogen, and upon the rate at which they equilibrate (Conrath; Gierasch 1984; Massie; Hunten 1982). For a simple and benchmark calculation, we assume that ortho- to para- ratio is fixed at 3:1 (normal hydrogen) and the heat capacities for other species are constant. The condensed phases are $NH_3(s)$, $H_2O(l)$, $H_2O(s)$, $NH_4SH(s)$, where "l" stands for liquid phase, and "s" stands for solid phase. The adiabatic temperature profile is generated to match a target temperature at 1 bar level, which is 166 K (Seiff et al. 1998). Figure 1 (a) shows the vertical profile of $NH_3$, $H_2O$ and $H_2S$. $H_2O$, $NH_4SH$ and $NH_3$ cloud layers form at 7.6 bar, 2.4 bar and 0.83 bar respectively. A small but visible kink near water mixing ratio equals $10^{-3}$ is due to triple point equilibrium. The increase of temperature due to freezing is recognized as a horizontal segment in the dry potential temperature profile in Figure 1 (b). Figure 1 (c) compares the numerical adiabatic lapse rate and its analytical value calculated by equation (28). They match exactly except for two places: one is at the triple point of water and the other is at the $NH_4SH$ cloud base. The analytic solution converges to the numerical solution at the wings near the triple point. Because the formation of $NH_4SH$ cloud does not satisfy the Clausius-Clapeyron relation, equation (28) cannot be applied to $NH_4SH$ condensation. For the approximate expression of lapse rate including $NH_4SH$ cloud, readers are referred to Atreya; Romani (1985).

Figure 2 shows the comparison with a traditional thermodynamic model (JAMRT) (Janssen et al. 2005), in which the thermal profile is obtained by integrating the moist adiabatic lapse rate. To compare, the JAMRT temperature profile is named as "integration profile". JAMRT has an adaptive mesh refinement scheme which inserts an additional grid point to identify the cloud bottom. Unlike our model, JAMRT uses constant latent heats for all condensable species. Three different latent heats for water are compared: dotted, dashed and solid blue lines represent $2.2\times10^6$, $2.38\times10^6$ and $2.5\times10^6$ J/kg respectively. They are the latent heats of water vapor at about 330K, 300K and 273K, covering the temperature range within which water condenses on Jupiter. The temperature difference between two completely different models are on the order of a fraction of a degree. However, the result of JAMRT is sensitive to the vertical resolution. A one-kilometer resolution model (shown in the green line in Figure 2) overestimates the effect of latent heat compared to the 100-m resolution model. The reason is that JAMRT integrates the moist adiabatic lapse rate from the bottom to the top using a quadrature rule. Because the moist adiabatic lapse rate as a function of height is concave up, i.e. having positive second derivative, the error is always negative and the quadrature rule overestimates the true value. Our method avoids the drawback by using iteration on equation (11) that guarantees convergence to machine precision for given values of entropy, pressure and abundance of each chemical species, independent of the mesh size. Moreover, the iterative method opens a simple and flexible way to calculate the secondary alteration of the atmosphere by dynamics or microphysics. For example, the next section introduces a stretch parameter that describes the subsidence of the atmosphere.

# 4. Interpretation of Galileo Probe result

The previous paragraphs studied an idealized model where the thermal profile is a moist adiabat. However, the real giant planet's atmosphere is far away from an idealized moist adiabat, as evidences from the recent 5 *μm* observation of Jupiter's atmosphere (Bjoraker et al. 2015), and as evidenced by the recent Juno microwave observations (Bolton et al. 2017; Li et al. 2017). In fact, studies with the Very Large Array (VLA) show depleted ammonia with respect to saturation for all four giant planets in the Solar System (de Pater; Massie 1985; de Pater et al. 2001). The same depletion of ammonia is also observed after Saturn's Giant Storm (Janssen et al. 2013; Laraia et al. 2013) and in the smaller storm in Saturn's southern hemisphere (Dyudina et al. 2007). Li; Ingersoll (2015) modeled the dynamic desiccation of ammonia after convection using numerical simulation, and they found that processes associated with geostrophic adjustment after convection deplete ammonia from saturation. Sugiyama et al. (2014) used a two-dimensional cloud-resolving model to show the explicit cycles of convective events. In their model, ammonia and water remain unsaturated during the quiescent period of the cycle. Since thermodynamics and dynamics are inevitably intertwined, and neither of them are understood well enough to give a conclusive picture of the atmosphere, here we give a simple parameterization for dynamic processes that modify the original moist adiabat.

Motivated by observations from the Galileo probe, by the numerical experiment that shows a downward deflection of material surfaces (vertical stretching of the air column) in Showman; Dowling (2000), and by an analytical wave saturation model by Friedson (2005), we simplify the dynamic distortion of the material surface to a scalar "stretch parameter" ($X$), so that the final pressure of the material surface ($p_2$) is $X$ times its original pressure ($p_1$) : $p_2 = Xp_1$. During the vertical stretch of the column, air parcels conserve their potential temperature and moisture contents. The stretch parameter effectively reduces the relative humidity of the atmosphere while maintaining the magnitude of stratification. We find that the vertical abundances of $NH_3$, $H_2S$ and $H_2O$ measured in situ by Galileo Probe are consistent with $X = 4$ (shown in Figure 3). Moreover, statically stable layers predicted by equilibrium condensation are preserved but displaced to higher pressures. In our stretched model, three stable layers occur at ~1.5 bars, ~7 bars and ~17 bars, which match the locations of stable layers at 0.5-1.7 bars, 3-8.5 bars, 14-20 bars derived by Magalhaes et al. (2002) from the $T$-sensor data of the Galileo probe. Because the value $X = 1$ gives an unaltered saturated moist adiabat, and $X = 4$ gives the observed mixing ratios of $NH_3$, $H_2S$, and $H_2O$ from the Galileo, by varying $X$, one can model any profile in between.

Modeling the hot spot by moving the pressure of the material surface seems to be an oversimplification of the dynamic processes. However, the fact that using one parameter is able to explain the profiles of all three condensates and the thermal stratification suggests some merits in the simple model. The key assumption embodied in the model is that material before the hot spot formed remains in the hot spot, and the material (air parcel) might plausibly have been saturated. Soundings inside several hurricane eyes on Earth have shown the enclosed air being drawn downward for about a few kilometers (Willoughby 1998), a weak evidence that supports the simple model. Therefore, the stretch model offers a plausible quantitative explanation for the Galileo observations. A detailed dynamic modeling in the future is of course needed to strengthen the argument.

# 5. Conclusion

In this work, we reviewed the published formulas and numerical methods for calculating moist adiabatic lapse rate in giant planet atmospheres. We derived a unified expression that holds for multiple condensable species, variable heat capacity and arbitrary amount of mixtures, and thus is applicable to both planets in the solar system and exotic exoplanets such as hot Jupiters and brown dwarfs. This expression reduces to the conventional moist adiabatic lapse rate when there is a single condensable species, and reduces to the Clausius-Clapeyron relation for a steam atmosphere. Moreover, we identified a cross term, $2(\eta_{NH_3}\beta_{NH_3})(\eta_{H_2O}\beta_{H_2O})$, that is missing in all published formulas for moist adiabats with multiple condensing species.

A numerical model is developed using the new formulation of thermodynamics, and validated by comparing the numerical adiabatic lapse rate with respect to the analytical one. We compared the thermal profile constructed by the new model with that constructed by a traditional thermodynamic model (JAMRT). The difference is on the order of a fraction of a degree.

Finally, we applied the new thermodynamic model to explain the Galileo Probe measurements. We introduced a stretch parameter ($X$) that describes the downwelling of a column of air. We found that the distribution of $NH_3$, $H_2S$ and $H_2O$ measured by Galileo Probe can all be fitted by $X = 4$, meaning that the air in the hotspot at 4 bar level is originated at 1 bar level. Using the stretch parameter, one can model any thermal and compositional profile ranging from the equilibrium condensation model to the Galileo Probe measurement.

# Acknowledgements

The research was carried out at Jet Propulsion Laboratory and California Institute of Technology. C.L. was supported by NASA Earth and Space Science Fellowship and by an internship at JPL working with Michael Janssen on the preparation for the Juno mission. We thank Michael Janssen's support for the internship and his kind assistance for hosting C.L. at JPL.

**Figure Caption List**

Figure 1 Standard Jupiter's troposphere constructed by iteration of equation (11). Panel (a): vertical distribution of $NH_3$ (green), $H_2S$ (magenta) and $H_2O$ (blue); their enrichment factors are 5, 3 and 5 respectively. Panel (b): temperature (dashed line, top axis) and dry potential temperature (solid line, bottom axis) profiles. The dry potential temperature is referenced at 1000 bar and is defined in equation (13). Panel (c): numerical and analytical adiabatic lapse rate. Blue dashed line is an approximation of adiabatic lapse rate by finite difference of the iteration temperature profile in panel (b). Red solid line is the analytical solution in equation (28).

Figure 2 Temperature profile difference after subtracting the integration profile from the iteration profile. Dotted, dashed and solid blues lines represent three choices of water latent heat, $2.2 \times 10^6$, $2.38 \times 10^6$ and $2.5 \times 10^6$ J/kg respectively, calculated at 100-meter resolution. The green line is calculated at one kilometer resolution when water latent heat is $2.5 \times 10^6$ J/kg.

Figure 3 Galileo probe results fitted by stretch parameter $X = 4$. Green lines represent $NH_3$ mixing ratio; blue lines represent $H_2O$ mixing ratio and magenta lines represent $H_2S$ mixing ratio. Dashed lines show the equilibrium condensation model with five times solar abundance for both $NH_3$ and $H_2O$. They do not match the Galileo probe results (Wong et al. 2004), which are the data points with error bars. The uppermost $NH_3$ point is an upper bound. Solid lines show the same amount of enrichment but with $X = 4$.

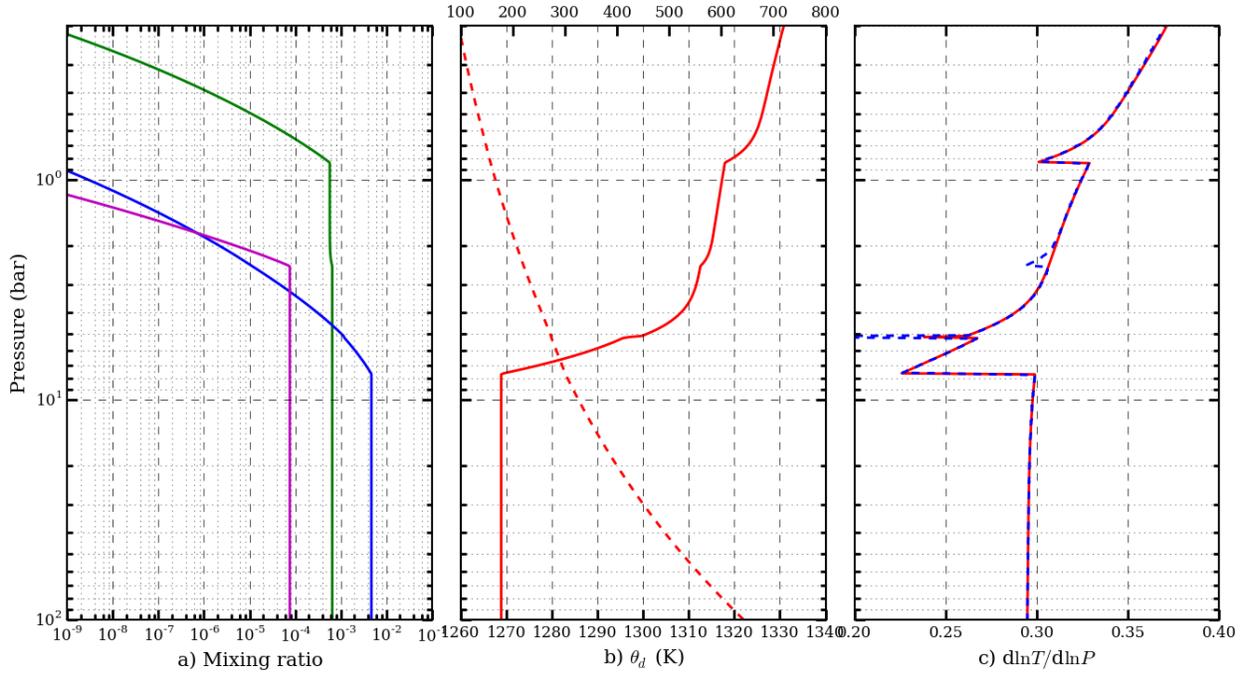

Figure 1 Standard Jupiter's troposphere constructed by iteration of equation (11). Panel (a): vertical distribution of $NH_3$ (green), $H_2S$ (magenta) and $H_2O$ (blue); their enrichment factors are 5, 3 and 5 respectively. Panel (b): temperature (dashed line, top axis) and dry potential temperature (solid line, bottom axis) profiles. The dry potential temperature is referenced at 1000 bar and is defined in equation (13). Panel (c): numerical and analytical adiabatic lapse rate. Blue dashed line is an approximation of adiabatic lapse rate by finite difference of the iteration temperature profile in panel (b). Red solid line is the analytical solution in equation (28).

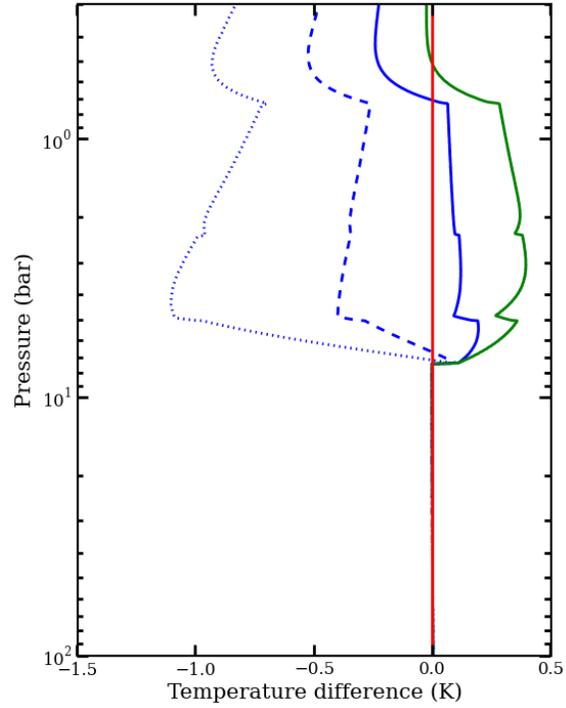

Figure 2 Temperature profile difference after subtracting the integration profile from the iteration profile. Dotted, dashed and solid blues lines represent three choices of water latent heat, $2.2\times10^6$, $2.38\times10^6$ and $2.5\times10^6$ J/kg respectively, calculated at 100-meter resolution. The green line is calculated at one kilometer resolution when water latent heat is $2.5\times10^6$ J/kg.

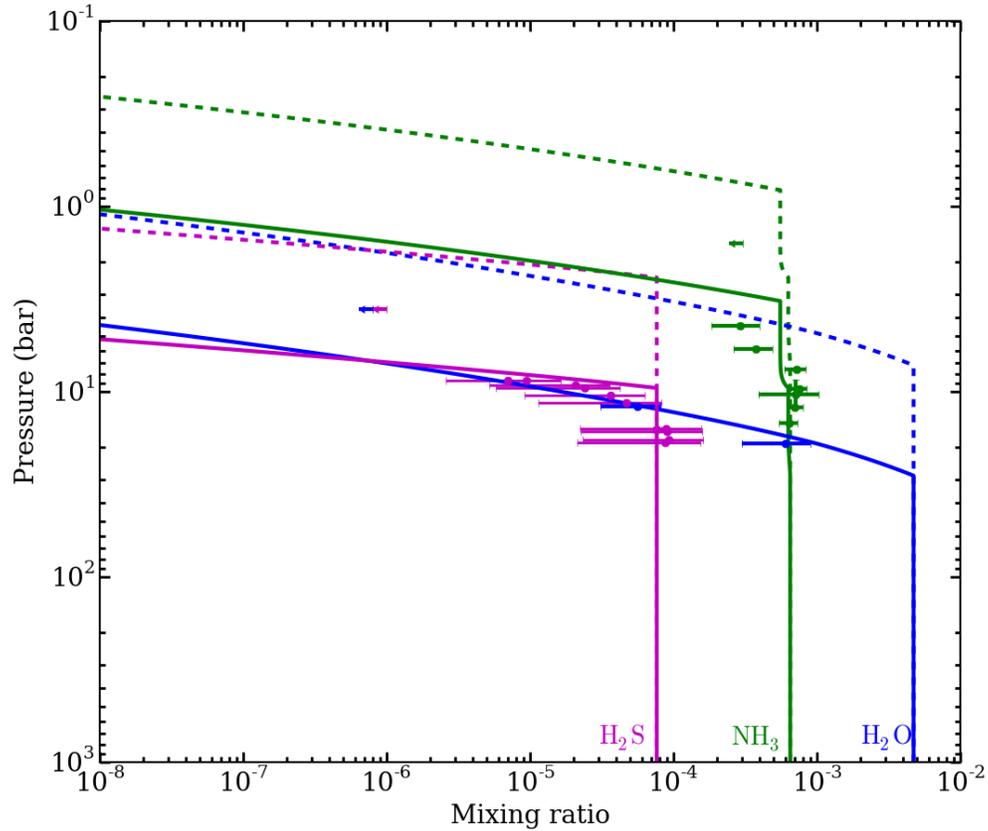

Figure 3 Galileo probe results fitted by stretch parameter $X = 4$. Green lines represent $NH_3$ mixing ratio; blue lines represent $H_2O$ mixing ratio and magenta lines represent $H_2S$ mixing ratio. Dashed lines show the equilibrium condensation model with five times solar abundance for both $NH_3$ and $H_2O$. They do not match the Galileo probe results (Wong et al. 2004), which are the data points with error bars. The uppermost $NH_3$ point is an upper bound. Solid lines show the same amount of enrichment but with $X = 4$.